# Kinematic Rotations of N-particle Hyperspherical Basis

L.L. Chachanidze-Margolin, *Marymount Manhattan College, New York.*

Sh. Tsiklauri, *Tbilisi State University, Georgia, Tbilisi.*

Abstract

The kinematical rotations of *N*-partical Hyperspherical basis are considered. The recurrence relation method of determination of transformation coefficients for arbitrary N-values is demonstrated.

## 1 Introduction

While investigating few-body systems with the use hyperspherical function method the problem of the hyperspherical basis transformation between different sets of Jacobi coordinates arises (the kinematic rotation problem) [1-6].
Coefficients of unitary transformations of three-particle hyperspherical functions under particle permutations were introduced in [1] (Raynal-Revai coefficients). For four- and more particle systems kinematic rotations include both particle permutations and transitions from one configuration to another. Four-particle transformations were discussed in [3,7-10].
The symmetrized basis may be constructed easily if the kinematic rotation coefficients are available [2,3,8,9,10].
The transformations of hyperspherical functions become sufficiently complex when the number of particles increases. The recurrence method [11,12] allows to avoid difficulties connected with the direct calculation of the kinematic-rotation coefficients.
A construction scheme for recurrence relations, proposed en [11], is applicable to the transformation coefficients for hyperspherical basis with any number of particles. First, according to this method the coefficients with $K_N = 2$, $L_N = 0$ with arbitrary hyperspherical function are expanded in terms of hyperspherical basis and the kinematical rotation of the expansion is performed with the use of the known coefficients with $K_N = 2$, $L_N = 0$. This procedure allows to obtain recurrence relations for the transformation coefficients with any $K_N$ and $L_N$. A general formula for the coefficients is not needed at all.
In [11] a way of obtaining transformation coefficients for $N = 3, 4, 5$ is shown. In the present paper the method is generalized for any number of particles.
In the next section the transformation coefficients with $K_N = 2$, $L_N = 0$ are considered.
In the third section the recurrence relations are obtained for N-body transformation coefficients.

## 2 Kinematic Rotations of N-particle Hyperspherical Basis at $K_N = 2, L_N = 0$.

The N-particle hyperspherical function is [13,14]:
$$\Psi_N^\tau(\Omega) = N_{K_N+\frac{3}{2}(n-2)}^{K_{N-1}+\frac{3}{2}(n-2),l_n}(\sin\alpha_{n-1})^{K_{N-1}}(\cos\alpha_{n-1})^{l_n}$$

$$p_m^{K_{N-1}+\frac{3}{2}n-\frac{5}{2};l_n+\frac{1}{2}}(\cos 2\alpha_{n-1})\left\{\Psi_{N-1}^{\bar\tau}(\overline\Omega)Y^{l_n}\left(\vec{\xi}_n\right)\right\}_M^L, \tag{1}$$

where $K_N$ are hypermoments of N-particle systems. $\Psi_{N-1}^{\bar\tau}(\overline\Omega)$ is $(N-1)$-particle hyperspherical function, $\bar\tau$ denotes the corresponding quantum numbers, $\vec{\xi}_n$ are Jacobi coordinates

$$n = N-1, \qquad \sin\alpha_{n-1} = \frac{\rho_{n-1}}{\rho_n},$$

$$\cos\alpha_{n-1} = \frac{\xi_n}{\rho_n}, \qquad 2m = K_N - K_{N-1} - l_n$$

$$\rho_v = \left(\sum_{\delta=1}^v \rho_\delta^2\right)^{\frac{1}{2}},$$

$$\Omega = \left(\alpha_1,...,\alpha_{n-1},\vec{\xi}_1,...,\vec{\xi}_n\right),$$

$$\overline\Omega = \left(\alpha_1,...,\alpha_{n-2},\vec{\xi}_1,...,\vec{\xi}_{n-1}\right),$$

$$N_\alpha^{b:c} = \sqrt{\frac{2d!(a+2)\Gamma(d+b+c+2)}{\Gamma(d+b+3/2)\Gamma(d+c+3/2)}}, \qquad 2d = a-b-c.$$

Let us perform a kinematic rotation of N-particle Jacobi vectors

$$\begin{vmatrix}\vec{\xi}_1\\\vec{\xi}_2\\....\\....\\\vec{\xi}_n\end{vmatrix} = \bar a \begin{vmatrix}\vec{\xi}_1\\\vec{\xi}_2\\....\\....\\\vec{\xi}_n\end{vmatrix} \tag{2}$$

Under the kinematic rotation (2) function (1) is transformed as follows
$$\Psi_N^\tau(\Omega) = \sum_{\tau'} <\tau'\mid\tau>_{K_N L_N}\Psi_N^{\tau'}(\Omega'), \tag{3}$$

where $\Omega, \Omega'$ are the sets of angles and hyperangles before and after transformation, $\tau, \tau'$ are corresponding quantum numbers.

For the kinematic rotation coefficients from eq. (3) we have
$$<\tau'\mid\tau>_{K_N L_N} = \int \Psi_N^{\tau'*}(\Omega')\Psi_N^\tau(\Omega)d\Omega. \tag{4}$$

In some cases the transformation coefficients may be found directly by solving integrals (4). It can be done, for example, at $K_N = 2, L_N = 0$, with taking into account that

$$(\sin\alpha_{n-1})^{K_{N=1}}p^{K_{N=1}+\frac{3}{2}N-4;\frac{1}{2}}(\cos 2\alpha_{n-1}) = \frac{3}{2}\left[(N-1)\xi_n^2 - \sum_{v=1}^n \xi_v^2\right]; \quad p_0 = 1. \tag{5}$$

Using eq. (5) in eq. (1), we have ($m = 0, 1, ..., n-2$)

$$\Phi^{l_1=0}_{K_{N-m}=2} = \varepsilon_{N-m}\frac{1}{\rho_n^2}\left[(n-m-1)\xi_{n-m}^2 - \sum_{v=1}^{n-m-1}\xi_v^2\right]; \quad \Phi^{l_1=l_k=1} = \varepsilon_0 \frac{1}{\rho_n^2}\vec{\xi}_i\vec{\xi}_k \qquad (6)$$

where $K_N = 2$ means that all $K_i(i < N)$ are equal to zero, while $K_{N-1} = 2$ means, that all $K_i(i < N-1)$ are equal to zero and so on. the condition $l_i = l_k = 1$ fixes all other quantum numbers at $K_N = 2$, $L_N = 0$.

From eqs (1,5,6) we can obtain the following relations for numberical coefficients

$$\frac{\varepsilon_{N-m}}{\varepsilon_{N-n}} = \sqrt{\frac{(N-n-1)(N-n-2)}{(N-m-1)(N-m-2)}}; \qquad (7)$$

$$\frac{\varepsilon_0}{\varepsilon_{N-m}} = -\sqrt{2(N-m-1)(N-m-2)}$$

With the use of eq. (6) we obtain

$$\frac{\xi_{n-m}^2}{\rho_n^2} = \frac{1}{n} - \sum_{f=1}^{m}\frac{1}{\varepsilon_{N-f+1}(n-f+1)(n-f)}\Phi_{K_{N-f+1}=2} + \frac{\Phi_{K_{N-m}=2}}{\varepsilon_{N-m}(n-m)} \qquad (8)$$

Using successively eqs (6,2,8,7) in eq. (4) we may obtain the kinematic rotation coefficients at $K_N = 2$, $L_N = 0$ as functions of elements $n_{ik}$ of the matrix (2) $(i,k = 1,...N-1)$.

$$\langle\Phi_{K_{N-f}=2}|\Phi_{K_{N-m}=2}\rangle_{20} =$$
$$\frac{(n-m-1)(n-f-1)a^2_{n-m,n-f} - (n-f-1)\sum_{i=m+1}^{n-1}a^2_{n-i,n-f}}{\sqrt{[(N-f-1)(N-f-2)(N-m-1)(N-m-2)]}} -$$
$$\frac{(n-m-1)\sum_{i=f+1}^{n-1}a^2_{n-m,n-i} - \sum_{i,i'=m+1}^{n-1}a^2_{n-i,n-i'}}{\sqrt{[(N-f-1)(N-f-2)(N-m-1)(N-m-2)]}};$$

$$\langle\Phi_{K_{N-m}=2}|l_i = l_k = 1\rangle_{20} = -\sqrt{\frac{2}{(N-m-1)(N-m-2)}} \qquad (9)$$

$$\left[a_{i,n-m}a_{k,n-m}(n-m-1) - \sum_{f=m+1}^{n-1}a_{i,n-f}a_{k,n-f}\right];$$

$$\langle l_i = l_k = 1|l_m = l_n = 1\rangle_{20} = a_{mi}a_{nk} + a_{mk}a_{ni};$$

$$\langle\Phi_{K_N=2}|\Phi_{K_N=2}\rangle_{20} = \frac{(N-1)a_{nm}^2 - 1}{N-2}.$$

In this case N-body coefficients (4) with $K_{N-1} = 2$ are the same functions of $a_{ik}$ $(i,k = 1,...,N-2)$ as the $(N-1)$-particle transformation matrix $\bar{a}$ isn't used

From eqs (2) and (4) it follows that

$$\langle\tau'|\tau\rangle_{K_N L_N}(a_{ik}) = \langle\tau|\tau'\rangle_{K_N L_N}(a_{ki}). \qquad (10)$$

## 3 Recurrence Relations for N-particle Kinematic Rotation Coefficients

Recurrence relations for hyperspherical basis transformation coefficients were obtained for three-, four- and five-body systems in [11]. Now we shall consider general principles of the recurrence relation construction for the N-particle hyperspherical basis

Let us rewrite formulae (6) using hyperangles $\alpha_i$ as follows:

$$\Phi_{K_{N-m}=2} = \bar{\varepsilon}_{N-m}\frac{1}{\rho_n^2}\left[\cos 2\alpha_{n-m-f} + \frac{N-m-3}{N-m-1}\right]\prod_{f=0}^{m-1}\sin\alpha_{n-m+f}^2; \qquad (11)$$

$$\Phi^{l_{n-1}=l_{n-k}=1} = \bar{\varepsilon}_0\frac{1}{\rho_n^2}\left\{Y^1\left(\vec{\xi}_{n-i}\right)Y^1\left(\vec{\xi}_{n-k}\right)\right\}_0^0 \cos\alpha_{n-i-1}\cos\alpha_{n-k-1}$$
$$\prod_{f=0}^{i-1}\sin\alpha_{n-i+f}\prod_{f'=0}^{k-1}\sin\alpha_{n-k+f'}; \qquad (12)$$

where $\bar{\varepsilon}_{N-m} = \frac{1}{2}\varepsilon_{N-m}(N - m - 1)$, $\bar{\varepsilon}_0 = -\frac{4\pi}{\sqrt{3}}\varepsilon_0$.

Using relations for Jacobi polynomials for products of functions (11,12) with an arbitrary hyperspherical function we obtain recurrence relations for hyperspherical functions. On the lefthand side of the relations we have products $\Phi_a \Psi_K^l$ ($a = m$ denotes functions (11), $a = i, k$ denotes functions (12), $\Psi_K^l$ is arbitrary hyperspherical function), but on the right-hand side we have linear combinations of functions $\Psi_K^l$ with changed quantum numbers, namely:

$$\Phi_a \Psi_K^l = \bar{\varepsilon}_a \sum_{\varepsilon \nu} Z_{\nu\varepsilon}^a \Psi_K^i(\nu, \varepsilon) \tag{13}$$

where $Z^a$ are numerical coefficients, $\varepsilon = 2, 0, -2$ changes $K_N(k_{N-\varepsilon} > K_{N+\varepsilon})$, and $\nu$ are parameters, determining changes of other $K$ and $L$.

Different functions of angles $\alpha_i$ in eqs (11,12) give rise to the following numerical coefficients and changes of $K, L$ on the right-hand side of eq. (13)

$$\left[\cos 2\alpha_{n-m-1} + \frac{N-m-3}{N-m-1}\right] \to$$
$$A_\xi \left(K_{N-m} + \frac{3}{2}(N - m - 3), l_{n-m}, K_{N-m-1} + \frac{3}{2}(N - m - 3)\right),$$
$$K_{N-m} \to K_{N-m} + \xi \quad (\xi = 2, -2, 0);$$

$$\sin \alpha_{n-i}^2 \to D^{\xi_1 \xi_2}\left(K_{N-i+1} + \frac{3}{2}(N - i - 3), K_{N-i} + \frac{3}{2}(N - i - 3), l_{n-i+1}\right),$$
$$K_{N-i+1} \to K_{N-i+1} + \xi_1, \ K_{N-i} \to K_{N-i} + \xi_2 \quad (\xi_1, \xi_2 = 2, -2, 0);$$

$$\cos \alpha_{n-i-1} \sin \alpha_{n-i-1} \to B_{\alpha\beta}^\xi \left(K_{N-i} + \frac{3}{2}(N - i - 3), l_{n-i}, K_{N-i-1} + \frac{3}{2}(N - i - 3)\right),$$
$$K_{N-i-1} \to K_{N-i-1} + \beta, \ l_{n-i} \to l_{n-i} + \alpha, \ K_{N-i} \to K_{N-i} + \xi,$$
$$\left(\xi = 2, -2, 0; \ \alpha, \beta = 1, -1\right); \tag{14}$$

$$\cos \alpha_{n-i-1} \to F_{\alpha\beta}\left(K_{N-i} + \frac{3}{2}(N - i - 4), l_{n-i+1}, K_{N-i-1} + \frac{3}{2}(N - i - 4)\right),$$
$$l_{n-i} \to l_{n-i} + \alpha, \ K_{N-i} \to K_{N-i} + \gamma \quad (\alpha, \gamma = 1, -1);$$

$$\sin \alpha_{n-i-1} \to F_{\alpha\gamma}\left(K_{N-i} + \frac{3}{2}(N - i - 4), K_{N-i-1} + \frac{3}{2}(N - i - 4), l_{n-i}\right)(-1)^{\frac{\alpha-\gamma}{2}};$$

$$\left\{Y^1\left(\vec{\xi_i}\right) Y^1\left(\vec{\xi_k}\right)\right\}_0^0 \to C_\alpha(l).$$

Coefficients $C_\alpha$ arise at the recoupling of spherical functions. For example, for $C$ with four indices, arising at $N = 5$, we have:

$$\left\{\left\{\left\{Y^1\left(\vec{\xi_1}\right) Y^1\left(\vec{\xi_4}\right)\right\}_0^0 \left\{Y^{l_1}\left(\vec{\xi_1}\right) Y^{l_2}\left(\vec{\xi_2}\right)\right\}^{L_3} Y^{l_3}\left(\vec{\xi_3}\right)\right\}^{L_4} Y^{l_4}\left(\vec{\xi_4}\right)\right\}_M^{L_5} =$$
$$\sum_{\alpha_1 \alpha_2 \alpha_3 \alpha_4} C_{\alpha_1 \alpha_2 \alpha_3 \alpha_4}(l_1, l_2, L_3 l_3, L_4, l_4, L_5) \tag{15}$$

$$\left\{\left\{\left\{Y^{l_1+\alpha_1}\left(\vec{\xi_1}\right) Y^{l_2}\left(\vec{\xi_2}\right)\right\}^{L_2+\alpha_2} Y^{l_3}\left(\vec{\xi_3}\right)\right\}^{L_4+\alpha_3} Y^{l_4+\alpha_4}\left(\vec{\xi_4}\right)\right\}_M^{L_5},$$
$$(\alpha_1, \alpha_4 = 1, -1; \ \alpha_2, \alpha_3 = 1, -1, 0)$$
$$\left\{Y^1\left(\vec{\xi_2}\right) Y^1\left(\vec{\xi_4}\right)\right\}_0^0 \to C_{\alpha_1 \alpha_2 \alpha_3 \alpha_4}(l_2, l_1, L_3 l_3, L_4, l_4, L_5).$$

It is clear from eq. (15) how to obtain $C_\alpha$ with more indices. For different concrete pairs $\xi_i, \xi_k$ we have

$$\left\{Y^1\left(\vec{\xi_1}\right) Y^1\left(\vec{\xi_2}\right)\right\}_0^0 \to C_{\alpha\beta}(l_1, l_2, L_3), \quad (\alpha, \beta = 1, -1),$$
$$\left\{Y^1\left(\vec{\xi_1}\right) Y^1\left(\vec{\xi_3}\right)\right\}_0^0 \to C_{\alpha\eta\beta}(l_1, l_2, L_3 l_3, L_4), \quad (\alpha, \beta = 1, -1; \ \eta = 1, 0, -1), \tag{16}$$

$$\{Y^1(\vec{\xi_2})Y^1(\vec{\xi_4})\}^0_0 \to (-1)^{\eta+1}C_{\alpha\eta\beta}(l_2,l_1,L_3,l_3,L_4).$$

Note that for systems with $N > 4$ the coefficients $A, B, C, D, F$ are the same as in the four-particle case, coefficients of spherical function recoupling $C_{\alpha_1...\alpha_{N-1}}$ must be added only.

For example, at $N = 6$, if $a = m = 2$ and if $a = i$, $k = 2,4$ eqs (13) are:

$$\Phi_{K_4=2}\Psi^i_k = \bar{\varepsilon}_4 \sum_{\tau\xi\varepsilon} A_\tau(K_4 + \tfrac{3}{2}, l_3, K_3 + \tfrac{3}{2})D^{\tau\xi}(K_5 + \tfrac{3}{2}, K_4 + \tfrac{3}{2}, l_4)$$
$$D^{\xi\varepsilon}(K_6 + 3, K_5 + 3, l_5)\Psi_{K_4+\tau,K_3+\xi,K_6+\varepsilon} \quad (\tau,\xi,\varepsilon = 2,0,-2),$$
(17)

$$\Phi_{l_2=l_4=l}\Psi^i_k = \bar{\varepsilon}_0 \sum_{\alpha\eta\zeta\beta\gamma\delta\xi\varepsilon} (-1)^{\frac{\gamma-\delta}{2}+\eta+1} F_{\alpha\gamma}(K_3, l_2, l_1)F_{\gamma\delta}(K_4 + \tfrac{3}{2}, K_3 + \tfrac{3}{2}, l_3)$$
$$C_{\alpha\eta\zeta\beta}(l_2,l_1,L_3,l_3,L_4,l_4,L_5)B^{\xi}_{\beta\delta}(K_5 + 3, l_4, K_4 + 3)D^{\xi\varepsilon}(K_6 + 3, K_5 + 3, l_5)$$
$$\Psi^{l_2+\alpha,L_3+\eta,L_4+\zeta,l_5+\beta}_{K_3+\gamma,K_4+\delta,K_5+\xi,K_6+\varepsilon}; \quad (\alpha,\beta,\gamma,\delta = 1,-1; \eta,\zeta = 1,0,-1; \xi,\varepsilon = 2,0,-2).$$

AS it was shown in [11,12], each of eqs (11),(12) leads to three ($\varepsilon = 2,-2,0$) recurrence relations for the transformation coefficients:

$$\sum_{v_0} Z^\alpha_{v_0\varepsilon}\langle \bar{l}, \bar{K}|lK(v)\rangle_{K_N+\varepsilon,L_N} = \sum_{\alpha'} \frac{\bar{\varepsilon}_{\alpha'}}{\varepsilon_\alpha}$$

$$\sum_v \overline{Z^\alpha_{v_0\varepsilon}}\langle \bar{l}, \bar{K}(-v)|lK\rangle_{K_N,L_N}\langle \alpha^{\bar{l}}|\alpha\rangle_{20},$$
(18)

where $\langle \alpha^{\bar{l}}|\alpha\rangle_{20}$ are coefficients (9) and

$$\frac{\bar{\varepsilon}_{N-m}}{\bar{\varepsilon}_{N-n}} = \sqrt{\frac{(N-m-1)(N-n-2)}{(N-n-1)(N-m-2)}};$$
$$\frac{\bar{\varepsilon}_0}{\bar{\varepsilon}_{N-m}} = 4\pi\sqrt{\frac{2(N-m-1)(N-m-2)}{3}}.$$
(19)

$\overline{Z^\alpha_{v\varepsilon}}$ is a coefficient $Z^\alpha_{v\varepsilon}$ with changed set of $K, l$ (in correspondence with $v$). For example, $Z^\alpha_{v\varepsilon}, \overline{Z^\alpha_{v\varepsilon}}$ at $N = 5$ $a = m = 2$ are:

$$Z^2_{\xi_1\xi_2\varepsilon} = A_{\varepsilon_1}(K_3, l_1, l_2)D^{\xi_1\xi_2}(K_4, K_3, l_3)D^{\xi_2\varepsilon}(K_5 + \tfrac{3}{2}, K_4 + \tfrac{3}{2}, l_4) \quad (20)$$
$$\overline{Z^2_{\xi_1\xi_2\varepsilon}} = A_{\varepsilon_1}(\bar{K}_3 - \xi_1, \bar{l}_1, \bar{l}_2)D^{\xi_1\xi_2}(\bar{K}_4 - \xi_2, \bar{K}_3 - \xi_1, \bar{l}_3)D^{\xi_2\varepsilon}(K_5 + \tfrac{3}{2}, \bar{K}_4 + \tfrac{3}{2}, \bar{l}_4).$$

The number of sets of recurrence relations is equal to the number of functions with $K_N = 2$, $L_N = 0$, namely:

$$\frac{(N+1)(N-2)}{2}.$$

At $K_N = L_N$ the formula for transformation coefficients for any number of particles was obtained in [15]. Starting with these coefficients one can use recurrence relations (18) for obtaining coefficients with any $K_N$ and $L_N$.

The expressions for coefficients $A, B, D, F$ and $C$ with 2, 3, 4, 5 indices are presented in the Appendix.

## 4 Appendix

$A, B, C, D, F$, coefficients entering recurrence relations for transformation coefficients.

$$A_2(K, l_1, l_2) = A_{-2}(K + 2, l_1, l_2) = \frac{1}{2(K+2)}$$
$$\left[\frac{(K-l_1-l_2+2)(K+l_1+l_2+4)(K+l_1-l_2+3)(K-l_1+l_2+3)}{(K+2)(K+4)}\right]^{\frac{1}{2}};$$

$$A_0(K, l_1, l_2) = \frac{(l_1-l_2)(l_1+l_2+1)}{(K+1)(K+3)} + \frac{N-3}{N-1};$$

$$B_{11}^2(K, l_1, l_2) = \frac{1}{4(K+3)} \left[ \frac{(K+l_1+l_2+4)(K+l_1+l_2+6)(K+l_1-l_2+3)(K-l_1+l_2+3)}{(K+2)(K+4)} \right]^{\frac{1}{2}};$$

$$B_{-1-1}^2(K, l_1, l_2) = -\frac{1}{4(K+3)} \left[ \frac{(K-l_1-l_2+2)(K-l_1-l_2+4)(K+l_1-l_2+3)(K-l_1+l_2+3)}{(K+2)(K+4)} \right]^{\frac{1}{2}};$$

$$B_{1-1}^2(K, l_1, l_2) = \frac{1}{4(K+3)} \left[ \frac{(K-l_1-l_2+2)(K+l_1+l_2+4)(K+l_1-l_2+3)(K+l_1-l_2+5)}{(K+2)(K+4)} \right]^{\frac{1}{2}};$$

$$B_{11}^0(K, l_1, l_2) = \frac{(l_2-l_1)\sqrt{(K+l_1+l_2+4)(K-l_1-l_2)}}{2(K+1)(K+3)};$$

$$B_{1-1}^0(K, l_1, l_2) = \frac{1}{2(K+1)(K+3)} (l_1 + l_2 + 1)[(K - l_1 + l_2 + 1)(K + l_1 - l_2 + 3)]^{\frac{1}{2}};$$

$$B_{\alpha\beta}^\nu(K, l_1, l_2) = (-1)^{\frac{\nu}{2}} B_{\beta\alpha}^\nu(K, l_2, l_1) = B_{-\alpha-\beta}^{-\nu}(K + \nu, l_1 + \alpha, l_2 + \beta);$$

$$C_{11}(l_1, l_2, L) = -\frac{\sqrt{3}}{8\pi} \left[ \frac{(L+l_1+l_2+2)(L+l_1+l_2+3)(l_1+l_2-L+1)(l_1+l_2-L+2)}{(2l_1+1)(2l_2+1)(2l_1+3)(2l_2+3)} \right]^{\frac{1}{2}};$$

$$C_{1-1}(l_1, l_2, L) = -\frac{\sqrt{3}}{8\pi} \left[ \frac{(L-l_1+l_2-1)(L-l_1+l_2)(l_1-l_2+L+1)(l_1-l_2+L+2)}{(2l_1+1)(2l_2+1)(2l_1+3)(2l_2-1)} \right]^{\frac{1}{2}};$$

$$C_{\alpha\beta}(l_1, l_2, L) = C_{\beta\alpha}(l_2, l_1, L) = C_{-\alpha-\beta}(l_1 + \alpha, l_2 + \beta, L);$$

$$F_{11}(K, l_1, l_2) = F_{11}(K - 1, l_1 + 1, l_2) = \frac{1}{2} \left[ \frac{(K-l_1-l_2) F_{11}(K-l_1+l_2+1)}{(K+1)(K+2)} \right]^{\frac{1}{2}};$$

$$D^{22}(K, K', l) = D^{-2-2}(K + 2, K' + 2, l) = \left[ \frac{(K-l+K'+8)(K-l+K'+6)(K+K'+l+7)(K+K'+l+9)}{(2K+7)(2K+9)^2(2K+11)} \right]^{\frac{1}{2}};$$

$$D^{2-2}(K, K', l) = D^{-22}(K - 2, K' + 2, l) = \left[ \frac{(K-l-K'+2)(K-l-K')(K-K'+l-1)(K-K'+l+1)}{(2K+3)(2K+5)^2(2K+7)} \right]^{\frac{1}{2}};$$

$$D^{20}(K, K', l) = -D^{-20}(K, K' + 2, l) = 2 \frac{[(K-l-K')(K-l+K'+6)(K+K'+l+7)(K-K'+l+1)]^{\frac{1}{2}}}{(2K+5)(2K+9)};$$

$$D^{02}(K, K', l) = D^{0-2}(K + 2, K', l) = -\left[ \frac{(K-l-K'+2)(K+l+K'+7)(K-K'+l+3)(K+K'-l+6)}{(2K+11)(2K+9)^2(2K+7)} \right]^{\frac{1}{2}};$$

$$D^{00}(K, K', l) =$$

$$\frac{(K-K'-l+4)(K-K'+l+5)(2K+9)+(K-K'-l+2)(K-K+l+3)(2K+5)}{(2K+5)(2K+7)(2K+9)};$$

$$C_{111}(l_1,l_2,l,l_3,L) = \frac{\sqrt{3}}{16\pi(l+1)}$$
$$\left[\frac{(L+l+l_3+2)(L+l+l_3+3)(l+l_3-L+1)(l+l_3-L+2)}{(2l_1+1)(2l_1+3)(2l_3+1)(2l_3+3)(2l+1)(2l+3)}\right]^{\frac{1}{2}}$$
$$[(l_1+l_2+l+2)(l_1+l_2+l+3)(l_1+l-l_2+1)(l_1+l-l_2+2)]^{\frac{1}{2}};$$

$$C_{-11-1}(l_1,l_2,l,l_3,L) = \frac{\sqrt{3}}{16\pi(l+1)}$$
$$\left[\frac{(L-l+l_3)(L-l+l_3-1)(l-l_3+L+1)(L-l-l_3+2)}{(2l_1+1)(2l_1-1)(2l_3+1)(2l_3-1)(2l+1)(2l+3)}\right]^{\frac{1}{2}}$$
$$[(l_1+l_2-l)(l_1+l_2-l-1)(l_2+l-l_1+1)(l_2+l-l_1+2)]^{\frac{1}{2}};$$

$$C_{11-1}(l_1,l_2,l,l_3,L) = -\frac{\sqrt{3}}{16\pi(l+1)}$$
$$\left[\frac{(L-l+l_3)(L+l-l_3+1)(l_3-l+L-1)(l-l_3+L+2)}{(2l_1+1)(2l_1+3)(2l_3-1)(2l_3+1)(2l+1)(2l+3)}\right]^{\frac{1}{2}}$$
$$[(l_1+l_2+l+2)(l_1+l_2+l+3)(l_1+l-l_2+1)(l_1+l-l_2+2)]^{\frac{1}{2}};$$

$$C_{-111}(l_1,l_2,l,l_3,L) = -\frac{\sqrt{3}}{16\pi(l+1)}$$
$$\left[\frac{(L+l+l_3+2)(L+l+l_3+3)(l+l_3-L+1)(l+l_3+L+2)}{(2l_1+1)(2l_1-1)(2l_3+1)(2l_3+3)(2l+1)(2l+3)}\right]^{\frac{1}{2}}$$
$$[(l_1+l_2-l)(l_1+l_2-l-1)(l_2+l-l_1+1)(l_2+l-l_1+2)]^{\frac{1}{2}};$$

$$C_{101}(l_1,l_2,l,l_3,L) = C_{10-1}(l_1,l_2,l,l_3+1,L) = -\frac{\sqrt{3}}{16\pi(l+1)l}$$
$$\left[\frac{(L+l+l_3+2)(l-L+l_3+1)(L+l_3-l+1)(l-l_3+L)}{(2l_1+1)(2l_1+3)(2l_3+1)(2l_3+3)}\right]^{\frac{1}{2}}$$
$$[(l_1+l_2+l+2)(l_1-l_2+l+1)(l_2+l-l_1+1)(l_2+l-l_1)]^{\frac{1}{2}};$$

$$C_{\alpha\beta\gamma}(l_1,l_2,l,l_3,L) = C_{-\alpha-\beta-\gamma}(l_1+\alpha,l_2,l+\beta,l_3+\gamma,L);$$

$$C_{\alpha\eta\varsigma\beta}(l_1,l_2,l,l_3,l',l_4,L) = \frac{1}{\sqrt{3}}m_\alpha(l_1)m_\beta(l_4)(-1)^{l_1+l_2+l_3+l_4+L}$$
$$[(2l_1+2\alpha+1)(2l+1)(2l+2\eta+1)(2l'+1)(2l'+2\zeta+1)(2l_4+2\beta+1)]^{\frac{1}{2}}$$
$$(-1)^{1+\eta}\begin{Bmatrix}l_2 & l_1+\alpha & l+\eta \\ 1 & l_1 & l\end{Bmatrix}\begin{Bmatrix}l_3 & l+\eta & l'+\varsigma \\ 1 & l' & l\end{Bmatrix}\begin{Bmatrix}L & l'+\varsigma & l_4+\beta \\ 1 & l_4 & l'\end{Bmatrix};$$

$$C_{\alpha\eta\varsigma\omega\beta}(l_1,l_2,l,l_3,l',l_4,l''l_5) =$$
$$\frac{1}{\sqrt{3}}m_\alpha(l_1)m_\beta(l_5)(-1)^{l_1+l_2+l_3+l_4+\eta+\varsigma+L}[(2l_1+2\alpha+1)(2l+1)(2l+2\eta+1)$$
$$(2l'+1)(2l'+2\zeta+1)(2l''+1)(2l''+2\omega+1)(2l_5+2\beta+1)]^{\frac{1}{2}}$$
$$\begin{Bmatrix}l_2 & l_1+\alpha & l+\eta \\ 1 & l & l_1\end{Bmatrix}\begin{Bmatrix}l_3 & l+\eta & l'+\varsigma \\ 1 & l' & l\end{Bmatrix}\begin{Bmatrix}l_4 & l'+\varsigma & l''+\omega \\ 1 & l'' & l'\end{Bmatrix}\begin{Bmatrix}L & l''+\omega & l_5+\beta \\ 1 & l_5 & l''\end{Bmatrix};$$

$$m_1(l) = \sqrt{\frac{3(l+1)}{4\pi(2l+3)}}; \quad m_{-1}(l) = -\sqrt{\frac{3l}{4\pi(2l-1)}}; \quad \{\ \} - 6j \text{ symbol}.$$

## References

[1] J.Raynal, J.Revai  Nuovo Cim. **68A** (1970) 612
[2] R.I.Jibuti, N.B.Krupennikova, V.Yu.Tomchinsky. Nucl.Phys. **A276** (1977) 421
[3] R.I.Jibuti, N.B.Krupennikova, L.L.Sarkisyan, Sh.M.Tsiklauri. Yad.Fiz. **44** (1986) 349

[4]  J.Macek, Kh.A.Teryian. Phys.Rev. **A33** (1986) 233
[5]  R.I.Jibuti, T.I.Efremidze, D.K.Tedoradze. Khim.Phys. **6** (1987) 34
[6]  J.M.Huston, S.Jain. J.Chem.Phys. **91** (1989) 4197
[7]  R.I.Jibuti, V.Yu.Tomchinsky, N.I.Shubitidze. Yad.Fiz. **18** (1973) 1164
[8]  R.I.Jibuti, N.B.Krupennikova, N.I.Shubitidze. Teor.Mat.Fiz. **32** (1977) 233
[9]  R.I.Jibuti, N.B.Krupennikova. *Hyperspherical-Function Method in Few-Body Quantum Mechanics*.Yad.Fiz. Tbilisi, Metsniereba, 1984
[10]  R.I.Jibuti, N.B.Krupennikova, L.L.Sarkisyan. Few-Body Systems,**4**, (1988) 151
[11]  N.B.Krupennikova. Yad.Fiz. **32** (1990) 940
[12]  R.I.Jibuti, N.B.Krupennikova, L.L.Chachanidze. Proc.X Europen Conference on Few-Body Physics. Uzhgorod, June 1-5 1990
[13]  N.Ya.Vilenkin, G.I.Kuznetsov, Ya.A.Smorodinsky. Yad.Fiz. **2** (1965) 906
[14]  M.S.Kildushov. Yad.Fiz. **15** (1972) 197
[15]  R.I.Jibuti, N.I.Shubitidze. Yad.Fiz. **32** (1980) 940